\documentclass{emulateapj}

\newcommand{\fig}[1]{Figure~\ref{#1}}

\newcommand{\name}{J121839.7+295325}
\newcommand{\zarc}{\ensuremath{2.48_{-0.05}^{+0.14}}}
\newcommand{\jlim}{\ensuremath{22.0}}
\newcommand{\hlim}{\ensuremath{20.7}}
\newcommand{\vlim}{\ensuremath{25.5}}

\begin{document}
\title{Is   the    Optically   Unidentified   Radio    Source,   FIRST
J121839.7+295325,  a  Dark  Lens?\footnote{Observations reported  here
were  obtained  at  the  MMT  Observatory, a  joint  facility  of  the
University of Arizona and the Smithsonian Institution}}
\shorttitle{FIRST \name}

\author{R. E.  Ryan Jr.\altaffilmark{2}, S.  H. Cohen\altaffilmark{3},
R. A.  Windhorst\altaffilmark{3,2}, C. R.  Keeton\altaffilmark{4}, and
T. J. Veach\altaffilmark{2}}

\altaffiltext{2}{Department  of  Physics,  Arizona  State  University,
Tempe,  AZ  85287-1504}  
\altaffiltext{3}{School  of Earth  and  Space
Exploration,   Arizona  State   University,  Tempe,   AZ,  85287-1404}
\altaffiltext{4}{Department   of   Physics   and  Astronomy,   Rutgers
University, Piscataway, NJ, 08854}

\shortauthors{Ryan Jr. et al.}

\email{russell.ryanjr@asu.edu}

\begin{abstract}

We  present evidence  that  the optically  unidentified radio  source,
FIRST J121839.7+295325,  may be strongly lensing  a background galaxy.
We estimate the redshift  of the assumed gravitational arc, discovered
in parallel imaging with HST, from MMT-Blue Channel spectroscopy to be
$z_{\rm  arc}\!=\!\zarc$.  We  present  lens models  with an  Einstein
radius    of   $R_E\!=\!1\farcs3$   which    contains   a    mass   of
$M_{dyn}\!=\!10^{12\pm0.5}$~M$_{\odot}$,    where    the   uncertainty
reflects the range  of possible lens redshifts.  The  putative lens is
not    detected   to    $J_{\rm   lim}\!=\!\jlim$~mag    and   $H_{\rm
lim}\!=\!\hlim$~mag in  our MMT-SWIRC imaging.  Using  the flux limits
from  WFPC2 and SWIRC,  we estimate  that the  dynamical mass-to-light
ratio               of               J121839.7+295325               is
$M_{dyn}/L_B\!\gtrsim\!10$~M$_{\odot}$~L$_{\odot}^{-1}$             for
$A_V\!=\!1$~mag,   and  this  lower   limit  could   be  as   high  as
$30$~M$_{\odot}$~L$_{\odot}^{-1}$  for   $A_V\!=\!0$~mag.   Since  the
radio  source is  optically unidentified  ($V_{\rm lim}\!=\!25.5$~mag)
and has a radio flux of $S_{1.4~{\rm GHz}}\!=\!33$~mJy, it is likely a
massive   early-type  galaxy   which   hosts  a   radio-loud  AGN   at
$0.8\!\lesssim\!z_{\rm  radio}\!\lesssim\!1.5$.  However,  the present
data cannot uniquely determine  the mass-to-light ratio of the lensing
galaxy, and hence the possibility that this system may be a reasonably
dark lens is not ruled out.

\end{abstract}


\section{Introduction} \label{intro}

In the $\Lambda$-dominated, cold dark matter ($\Lambda$CDM) cosmology,
massive  galaxies form  hierarchically  inside of  dark matter  haloes
\citep[eg.][]{wf91,kauf93}.   The mass evolution  of these  CDM haloes
can   place   stringent  constraints   on   the  cosmological   models
\citep[eg.][]{gre02}, and  the most reliable measurements  of CDM halo
masses  are  generally derived  from  modeling gravitationally  lensed
images. Furthermore, gravitational lensing  has become a powerful tool
for  studying properties  of both  the  lensing objects  and the  more
distant sources.   Of particular interest are constraints  on the mass
profiles  \citep[eg.][]{koop06,more}   and  the  mass-to-light  ratios
\citep[eg.][]{kkf98,rk05,tr06} of lens galaxies, and the host galaxies
of active  galactic nuclei \citep[eg.][]{kkm01,peng06}.   While all of
these studies are advancing due  to increased numbers of known lensing
systems   \citep{rat99,leh00,bol06},  they   are   still  limited   by
relatively small sample sizes.

In  this work,  we report  the discovery  of an  optical arc  which is
$\sim\!4\farcs0$ southwest of the optically unidentified radio source,
FIRST \name.  This radio source  was discovered in the Faint Images of
the Radio Sky at Twenty-centimeters  (FIRST) with the Very Large Array
\citep[VLA;][]{beck,white} and later  detected at 610~MHz \citep{reng}
and  74~MHz  \citep{cohen}.  Its  relatively  bright  flux of  $S_{\rm
1.4~{\rm   GHz}}\!=\!33$~mJy   suggests   that   it  is   at   $z_{\rm
radio}\!\lesssim\!1.5$    \citep[eg.][]{deb01,wad01}.    Additionally,
\name\  was  \textit{undetected}  in  the  pure-parallel  observations
(PropID:~8090,  PI: S.~Casertano)  with the  Wide Field  and Planetary
Camera~2   (WFPC2)   on   the   Hubble  Space   Telescope   (HST)   to
F606W$\gtrsim\!25.5$~mag  \citep[$V$-band  hereafter;][]{jqr07}, which
indicates    it   is    likely   at    $z_{\rm   radio}\!\gtrsim\!0.8$
\citep[eg.][]{win85,kron85,deb02,win03}.     These   two   independent
constraints yield a coarse redshift estimate of $0.8\!\lesssim\!z_{\rm
radio}\!\lesssim\!1.5$.   By correlating  the WFPC2  archive  with the
catalog of FIRST sources,  \citet{jqr07} note an arc-like feature with
$V_{\rm   Vega}\!=\!24.0\pm0.1$~mag.    Based   on   the   astrometric
uncertainties for the HST-WFPC2  and FIRST imaging, they conclude that
the  likelihood that  this arc  is the  optical identification  of the
radio source is ${\cal  L}\!\lesssim\!10^{-10}$, following the work of
\citet{der}.  Since  \citet{jqr07} rule  out the possibility  that the
arc is  the optical identification  of \name, we will  investigate the
hypothesis that the optical arc  is a gravitationally lensed imaged by
the optically unidentified radio source.

This  work is  organized as  follows: \S~\ref{mmt}  describes  our MMT
observations,   \S~\ref{model}    outlines   the   lensing   analysis,
\S~\ref{discuss}  discusses   the  lensing  interpretation   of  these
observations,  and  \S~\ref{con}   gives  some  closing  remarks  with
thoughts  toward  possible  future  observations.   Unless  explicitly
stated, all magnitudes  are in the AB system  \citep{abmag}.  We adopt
the     following     cosmological     model:     $\Omega_0\!=\!0.24$,
$\Omega_{\Lambda}\!=\!0.76$, and $H_0\!=\!100h$~km~s$^{-1}$~Mpc$^{-1}$
where $h\!=\!0.73$ \citep{wmap}.

\section{The MMT Observations}\label{mmt}

\subsection{Blue Channel Spectroscopy} \label{bcspec}

While  the  radio and  optical  properties  provide  a broad  redshift
constraint   on   the   lensing   object   of   $0.8\!\lesssim\!z_{\rm
radio}\!\lesssim\!1.5$,  little  is   known  about  the  optical  arc.
Therefore, we observed the putative  gravitational arc on May 18, 2007
with  the blue-channel  spectrograph on  the 6.5~m  MMT,  and obtained
4$\times$2400~s    and   1$\times$1200~s    exposures.     For   these
observations, our  $1\arcsec\!\times\!180\arcsec$ slit tracked through
the sky at  the parallactic angle, as the object  varied in airmass of
$\sec z\!\leq\!1.17$, which was very closely aligned with the position
angle  of the  arc seen  in  the left  panel of  \fig{mod}.  With  the
500~grooves~mm$^{-1}$  grating, the  spectral resolution  was 3.6~\AA\
with  a  dispersion   of  1.2~\AA~pix$^{-1}$.   We  observed  Feige~34
throughout the  night to  determine the spatial  point-spread function
(PSF) and to serve as the spectrophotometric flux standard star.

The two dimensional spectra  were reduced by standard algorithms using
custom  routines in IDL\footnote{http://www.ittvis.com/idl}.   The CCD
read  noise,  gain, and  dark  rate  were ${\cal  R}\!=\!2.5$~$e^{-}$,
${\cal        G}\!=\!1.1$~$e^{-}$~adu$^{-1}$,        and        ${\cal
D}\!=\!28.7$~$e^{-}$~pix$^{-1}$~hr$^{-1}$, respectively.  To determine
an accurate trace, we simply  rebinned the two dimensional spectrum in
wavelength to  boost the  relatively poor signal-to-noise  ratio.  The
one dimensional  spectrum was extracted  from a given frame  using the
observed  PSF  as  weights  in  both axes.   Finally,  the  wavelength
solution for each frame was determined from an HeNeAr comparison lamp,
which was observed at the corresponding position on the sky.

Despite  the long  integration time  (3~hours) on  the arc,  the total
signal-to-noise per  pixel is relatively low.   Therefore, we rebinned
the    calibrated,   sky-subtracted,    and   stacked    spectrum   to
$\Delta\lambda\!=\!125$~\AA\     for    $\lambda\!<\!5400$~\AA\    and
$\Delta\lambda\!=\!375$~\AA\  for $\lambda\!\geq\!5400$~\AA.   At this
resolution,    a   weak    spectral   break    is    identifiable   at
$\lambda\!\simeq\!4300$~\AA, however  we cannot conclusively determine
whether  it is  the Lyman  ($\lambda_{\rm rest}\!=\!1216$~\AA)  or the
Balmer/4000~\AA\  break.   Therefore,  to  determine the  most  likely
redshift with its associated  uncertainty and relative probability, we
computed a spectrophotometric  redshift \citep[eg.][]{ryan07} with the
redshift code, \texttt{HyperZ} \citep{hyperz}.  We did not include the
$V$-band flux in  the fit, since the MMT slit  did not fully encompass
the  HST-WFPC2  aperture.  We  adopted  a  grid  of synthetic  spectra
\citep{bc03}  parameterized by an  exponential star  formation history
with $e$-folding times of $\tau\!=\!0,~1,~2,~3,~5,~15,~30,~\infty$~Gyr
and ages  of $1~{\rm  Myr}\!\leq\!t\!\leq\!T(z)$, where $T(z)$  is the
age of the Universe.  The  final grid of spectral templates is defined
by the extinction of $A_V\!\leq\!3~{\rm mag}$ for $z\!\leq\!6$.

\begin{figure}
\epsscale{1.0}
\plotone{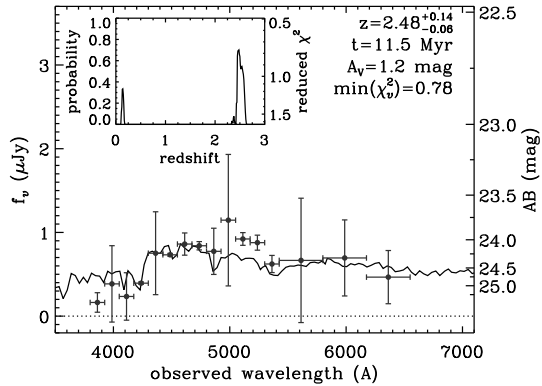}
\caption{The  observed  MMT spectrum  of  the  arc.   To increase  the
signal-to-noise to  a usable level,  we resampled the  one dimensional
spectrum          to          $\Delta\lambda\!=\!125$~\AA\         for
$\lambda\!\leq\!5400$~\AA\    and   $\Delta\lambda\!=\!375$~\AA\   for
$\lambda\!>\!5400$~\AA,  which is  shown as  filled circles.   Since a
weak spectral break  is readily apparent at $\lambda\!\sim\!4300$~\AA\
with no other noticeable  spectral features, we determine the redshift
from the photometric redshift code \texttt{HyperZ} \citep{hyperz}.  In
the inset, we show the  relative probability and reduced $\chi^2$ used
to determine  the redshift. In the  upper right, we  give the relevant
\texttt{HyperZ} fitted parameters.}\label{spec}
\end{figure}

In \fig{spec}, we show the observed spectrum as filled points with the
best fit template as a solid line.  In the inset, we show the relative
probability,  which  indicates two  possible  redshift solutions:  the
Balmer/4000~\AA\  break  at  $z\!\sim\!0.13$  or the  Lyman  break  at
$z\!\sim\!2.5$.  If the arc-like object is at $z\!\sim\!0.13$, then it
is   likely   an   unrelated   foreground  galaxy   to   \name\   with
$M_V\!\sim\!-14$~mag.  However, if it  is at the higher redshift, then
the  observed arc morphology  in the  HST-WFPC2 imaging  suggests that
\name\ may be gravitationally  lensing a background source.  Without a
flux measurement blueward of  the blue channel spectrum ($\lambda_{\rm
obs}\!\lesssim\!3800$~\AA)   to   break   this   redshift   degeneracy
\citep[eg.][]{hyperz},  we   can  only  make   statistical  statements
regarding the redshift of the  arc and the physical interpretations of
the   system.   From   the   inset  in   \fig{spec},  the   integrated
probabilities  of   the  two   peaks  are  13\%   and  87\%   for  the
Balmer/4000~\AA\ and Lyman breaks, respectively. Since the Lyman break
redshift is $\sim\!6.7$ times more likely, we will further develop the
lensing  hypothesis.  The best-fit  template found  by \texttt{HyperZ}
has   an   age   of   $t\!=\!11.5$~Myr,   instantaneous   burst   with
$A_V\!=\!1.2$~mag,  which is similar  to the  Lyman break  galaxies at
$2\!\lesssim\!z\!\lesssim\!3.5$  \citep{pap01}.   Therefore  the  most
likely  redshift of arc  is $z_{\rm  arc}\!=\!\zarc$, which  we derive
from $\chi^2(z)\!=\!\min{[\chi^2(z)]}$ and $\Delta\chi^2(z)\!=\!1$.

\subsection{SWIRC Infrared Imaging} \label{swirc}

We observed \name\ with the  SAO Wide-field InfraRed Camera (SWIRC) on
March  19,  2008  and  obtained  $80\times120$~s  and  $245\times30$~s
dithered images  in the $J$- and $H$-bands,  respectively.  These data
were  reduced by  standard  algorithms  with a  similar  suite of  IDL
routines  used  for  the  blue-channel  reductions.   The  seeing  was
relatively  constant   through  the  night   at  $\sim\!0\farcs8$  and
$\sim\!0\farcs6$  in the  $J$- and  $H$-bands, respectively.   The sky
image   per  object  frame   was  created   using  \texttt{SExtractor}
\citep{sex}.   In the  $5\farcm12\times5\farcm12$  field-of-view, each
object frame contained three stars  from the Two Micron All Sky Survey
(2MASS)  point  source catalog  \citep{cut,skrut},  which  we used  to
calibrate  the astrometry  of each  frame  and to  determine the  flux
zeropoint, with the  ($m_{\rm Vega}-m_{\rm AB}$) magnitude conversions
of \citet{rud}.  The  astrometrically corrected, sky-subtracted images
were  averaged together  with  a 1$\sigma$-clip  into stacked  science
frames.

We  disregard a  small portion  of  the SWIRC  field-of-view owing  to
appreciable  cosmetic  defects.  Therefore,  the  stacked images  each
contain  $\sim\!100$~objects   and  yield  number   counts  which  are
consistent with the  $J$-band galaxy counts of the  Chandra Deep Field
and  Hubble  Deep Field,  South  \citep[][see  their Figure  2]{sara}.
Based  on a  power-law fit  to our  observed counts,  we  estimate the
$3\sigma$,  50\% completeness  limits are  $J_{50\%}\!=\!21.0$~mag and
$H_{50\%}\!=\!19.6$~mag.  The assumed lensing galaxies is not detected
in either of these bandpasses, which may suggest that \name\ is indeed
a  rather  dark object.   To  estimate  our  detection limits  in  the
infrared imaging, we determine  the limit corresponding to a $3\sigma$
fluctuation  at the  position of  the  radio source.   For a  circular
aperture  with  a radius  of  $3\times{\rm  FWHM}$,  these limits  are
$J_{\rm lim}\!=\!\jlim$~mag and $H_{\rm lim}\!=\!\hlim$~mag and are 
used below to derive the limits on the mass-to-light ratios.

\section{The Gravitational Lens Model}\label{model}

If we  are to  interpret this system  in the context  of gravitational
lensing, we  must investigate whether  lensing can provide  a sensible
explanation of the optical arc.  The morphology of the arc immediately
implies that  the lensing galaxy  cannot be circular: a  circular lens
that produces such a long  arc (compared with the radius of curvature)
would  produce  a prominent  counter-arc  as  well.   In contrast,  an
elongated  galaxy  can   produce  a  single  long  arc   either  as  a
highly-distorted single image, or as a merged triplet of images.  That
leaves  the question  of how  to interpret  the compact,  high surface
brightness knot just  off the East end of the arc.   Since there is no
natural way to  interpret the arc and knot as  lensed images of single
source, the  knot is either a lensed  image of a second  source, or an
unrelated  foreground object.   We attempted  to model  the knot  as a
lensed image, although  it has little impact on  the results since the
observed image merely determines  the properties of the assumed second
source.

\begin{figure*}\centerline{\epsfxsize=0.75\hsize{\epsfbox{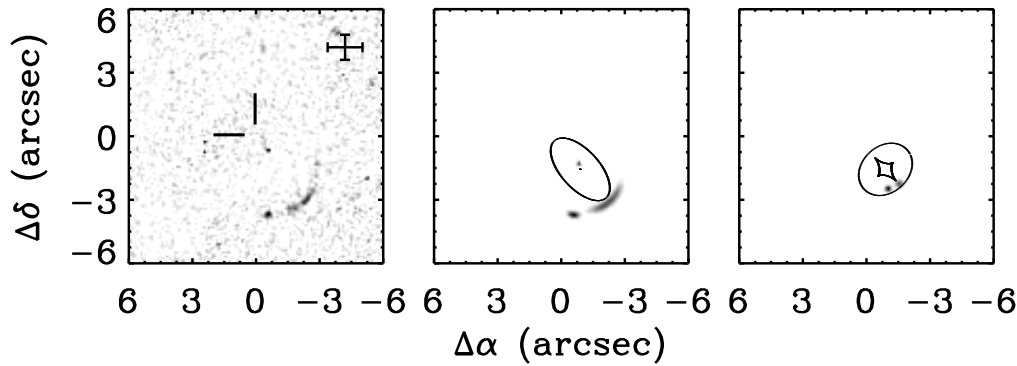}}}\vspace*{0.1in}
\caption{The  gravitational   lens  model.   We   show  the  HST-WFPC2
observations with the radio position indicated by the tick marks (left
panel)  and   the  1$\sigma$  astrometric   uncertainties,  the  model
reconstruction  (middle  panel), and  the  source plane  configuration
(right panel).   We show the critical  curves and the  caustics on the
model and source plane, respectively. We model the source plane as two
Gaussians and minimize the $\chi^2$  between the WFPC2 imaging and the
lens    model   to    determine   their    positions,    fluxes,   and
sizes.}\label{mod}
\end{figure*}

\begin{figure}
\epsscale{1.0}
\plotone{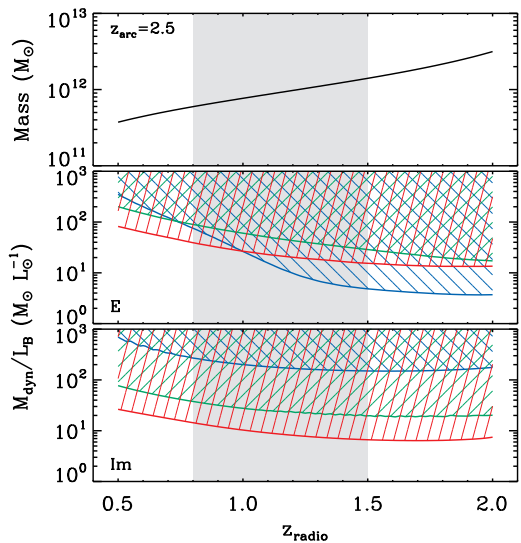}
\caption{The lens properties.   In the upper panel, we  show the total
mass enclosed  by the Einstein  radius of $R_E\!=\!1\farcs3$.   In the
middle  and  lower  panels,  we  show  the  limits  on  the  dynamical
mass-to-light ratio in the rest-frame $B$-band ($M_{dyn}/L_B$) derived
from     the    $V_{\rm     lim}\!=\!\vlim$~mag     (blue),    $J_{\rm
lim}\!=\!\jlim$~mag  (green),  and  $H_{\rm lim}\!=\!\hlim$~mag  (red)
observed flux limits  for $A_V\!=\!0$~mag.  We show the  range of most
probable    lens    redshift    with    the   shaded    region    (see
\S~\ref{intro}).}\label{mass}
\end{figure}

We did not  make a prior assumption about whether the  arc is a single
image or merged triplet.  Instead, we merely postulated that there are
two sources behind  an elongated lens galaxy.  We  treated the sources
as  circular Gaussians  for simplicity,  fitting for  their positions,
sizes,  and fluxes.   We  treated  the lens  galaxy  as an  isothermal
ellipsoid, fitting for its position, Einstein radius, ellipticity, and
position  angle.  For a  given set  of model  parameters, we  used the
software package  \texttt{lensmodel} \citep{kee01} to  find the images
of the two sources, which we  compared to the observed arc and knot on
a pixel-by-pixel  basis to  compute the $\chi^2$  goodness-of-fit.  We
used a downhill simplex method  to search the parameter space, with an
inner loop to optimize the  source parameters for a fixed lens galaxy,
and an outer  loop to optimize the galaxy parameters  as well.  In the
best-fit model, the lens  has an Einstein radius of $R_E\!=\!1\farcs3$
and  ellipticity of  $e\!=\!0.5$,  the two  sources  are separated  by
$\Delta\theta\!=\!0\farcs6$,   and    the   reduced-chi   squared   is
$\min{[\chi^2_{\nu}]}\!=\!0.94$.  In \fig{mod},  we show the HST-WFPC2
image (left),  the lens model  (middle), and the  reconstructed source
plane (right).   The mass enclosed  within the Einstein  radius slowly
varies        with       the        lens        redshift       between
$M_{dyn}\!=\!10^{12.5\pm0.5}$~M$_{\odot}$  (see  the  upper  panel  of
\fig{mass}), which is a relatively  large mass to have been undetected
in the HST-WFPC2 and MMT-SWIRC imaging.

There are three  additional points to make regarding  the lens models.
(1) In successful  models, the arc is a  single highly-distorted image
of a source lying just outside a cusp caustic in the source plane.  We
did not  impose this as an  assumption, it emerged  naturally from the
modeling.  (2)  The present  model does not  explain the  variation in
surface brightness  along the arc,  because we used a  simple Gaussian
source simply to reproduce the shape of the arc.  If the arc is indeed
singly-imaged,  then the  surface brightness  variation along  the arc
merely  represents the intrinsic  variations of  the source.   (3) Our
model predicts  a compact, faint counter-image just  beyond the center
of the lensing galaxy (see  middle panel of \fig{mod}).  There appears
to be some flux in roughly  the right location in the HST-WFPC2 image,
however it is likely associated with  a hot pixel.  We did not use the
portion of the image as constraints to the lens models.

Since \name\  is undetected in  the HST and  MMT imaging, we  can only
determine a lower  limit on its mass-to-light ratio.   We must convert
the above  detection limits ($J_{\rm  lim}$ and $H_{\rm lim}$)  to the
luminosity  in a fixed  bandpass to  determine a  mass-to-light ratio,
which is  independent of  redshift.  To facilitate  direct comparisons
with  other galaxies,  we  opt to  $k$-correct  our limiting  apparent
magnitudes  to  the rest-frame  $B$-band  absolute magnitudes.   These
conversions require assuming a template SED for the undetected lensing
galaxy \citep[eg.][]{hogg}.   To highlight two  extreme possibilities,
we show the range of the dynamical mass-to-light ratio consistent with
the $V$-band  (blue), $J$-band (green), and $H$-band  (red) imaging as
derived from the E (middle) and Im (bottom) templates \citep{cww80} in
\fig{mass}.   The shaded  region shows  the likely  lens  redshift, as
discussed in \S~\ref{intro}.  If a stellar component to \name\ exists,
then we  expect it likely  has an early-type  based on its  radio flux
\citep[eg.][]{win85}.    Therefore,   to   be  compatible   with   the
observations,              \name\              likely              has
$M_{dyn}/L_B\!\gtrsim\!30$~M$_{\odot}$~L$_{\odot}^{-1}$,  and could be
as  dark  as  $M_{dyn}/L_B\!\gtrsim\!150$~M$_{\odot}$~L$_{\odot}^{-1}$
for the  late-type SED.  These limits  will slightly decrease/brighten
by  including dust  obscuration, since  some fraction  of  the stellar
light may be extincted and  increase the maximum luminosity, for fixed
flux  limits.   However,  even  with  a modest  amount  of  extinction
($A_V\!=\!1.0$~mag), the intrinsic  mass-to-light ratio would still be
$M_{dyn}/L_B\!\gtrsim\!10$~M$_{\odot}$~L$_{\odot}^{-1}$    (for    the
early-type template).   Nevertheless, the WFPC2  $V$-band observations
generally provide  the strongest  constraint, owing to  the relatively
shallow SWIRC observations.

\section{Discussion} \label{discuss}

The HST-WFPC2 imaging suggests  that \name\ is gravitationally lensing
a  background object.   Our  blue channel  spectroscopy supports  this
hypothesis  by placing  the optical  arc  is likely  at high  redshift
($z_{\rm arc}\!=\!\zarc$).   Given the flux limits  from HST-WFPC2 and
MMT-SWIRC  ($V_{\rm lim}\!=\!\vlim$~mag,  $J_{\rm lim}\!=\!\jlim$~mag,
and $H_{\rm  lim}\!=\!\hlim$~mag), we find  the lensing object  may be
rather    dark,    with   a    $B$-band    mass-to-light   ratio    of
$M_{dyn}/L_B\!\gtrsim\!30$~M$_{\odot}$~L$_{\odot}^{-1}$.  From the K20
survey,   \citet{sper}  find  a   dynamical  mass-to-light   ratio  of
$M_{dyn}/L_B\!\lesssim\!2$~M$_{\odot}$~L$_{\odot}^{-1}$    for   their
sample of 18~early-type galaxies  at $z\!\simeq\!1$. Their upper limit
can                   be                  extended                  to
$M_{dyn}/L_B\!\lesssim\!8$~M$_{\odot}$~L$_{\odot}^{-1}$ by including a
sample  of  28~galaxies  from  the Coma  cluster  \citep{guit}.   This
mass-to-light  ratio is roughly  consistent with  the $z\!\simeq\!0.5$
early-type      galaxies      of      \citet{vnd},     which      have
$2.5\!\lesssim\!M_{dyn}/L_B\!\lesssim\!6.8$~M$_{\odot}$~L$_{\odot}^{-1}$.
By  comparing to  a series  of $z\!\simeq\!0$  studies, they  find the
mass-to-light    ratio    of    early-type   galaxies    evolves    as
$\Delta\log{(M_{dyn}/L_B)}/\Delta             z\!=\!-0.457\pm0.046~(\rm
random)\pm0.78~(\rm systematic)$,  since the low  redshift ellipticals
have           $M_{dyn}/L_B\!\lesssim\!15$~M$_{\odot}$~L$_{\odot}^{-1}$
\citep{pad}.  Therefore, if our lower limits derived in \S~\ref{model}
are correct, then \name\ may  be a relatively dark galaxy, whose total
mass is dominated  by dark matter.  While there  are solid statistical
arguments that such objects should exist \citep{ps74,bla01}, there are
currently no confirmed examples.  Since direct optical observations of
these extreme  mass-to-light objects  will be nearly  impossible, they
will likely manifest themselves as ``dark lenses'' \citep{rus02}.

The  most  common  suggestions   of  dark  lensing  have  arisen  from
wide-separation  quasar pairs (WSQPs),  quasars with  nearly identical
optical               spectra               separated               by
$3\arcsec\!\lesssim\!\Delta\theta\!\lesssim\!10\arcsec$,    with    no
identifiable  lensing source.   Detailed  analyses of  the spectra  of
these quasars, from the radio  to the $X$-rays, have revealed that all
known        WSQPs       are        simply        binary       quasars
\citep{haw97,mun98,mor99,peng99,mor00,gre02}.     Additionally,    the
Cosmic Shear  survey with HST  \citep{css} has identified a  number of
gravitational   lenses.   In   one  particular   field,  \citet{mir02}
suggested that  the conspicuous tangential alignment  of galaxies with
arc-like  morphologies  and  no  obvious  over-density  of  foreground
galaxies may  be a dark  lens.  However, \citet{erb03} ruled  out this
hypothesis  by   demonstrating  the  colors  of   these  galaxies  are
inconsistent with high redshift, lensed objects.

Since \name\ remains unidentified in the optical and near-infrared, it
is  a good  candidate  for a  dark  lens.  However,  the bright  radio
emission  of  $S_{1.4~{\rm GHz}}\!=\!33$~mJy  suggests  that the  lens
object     likely    contains     an    active     galactic    nucleus
\citep[eg.][]{win85}.  If a baryonic  counterpart to this radio source
exists, then it likely corresponds  to a massive and perhaps extincted
galaxy,  which would  make  it  similar to  the  distant red  galaxies
\citep[DRGs;][]{fs04}     or     the     extremely     red     objects
\citep[EROs;][]{eros}.   If we  assume the  optical/infrared  color of
$(R-K)\!\gtrsim\!5.5$~mag  typical of  an ERO,  then \name\  must have
$K\!\gtrsim\!20$~mag       and      a       stellar       mass      of
$M_{*}\!\lesssim\!10^{10}$~M$_{\odot}$    \citep[][based    on   their
Figure~7]{cons08}.  In  this case, our  lens models imply  that \name\
would     have     a     dynamical-to-stellar    mass     ratio     of
$M_{dyn}/M_{*}\!\gtrsim\!100$, which is  significantly higher than the
$z\!\simeq\!2$  star-forming galaxies  with $M_{dyn}/M_{*}\!\simeq\!2$
\citep{erb}.

\section{Conclusion} \label{con}
The HST-WFPC2  and MMT data suggest  that \name\ is  an extremely dark
galaxy  worthy of follow-up  observations to  confirm or  exclude this
hypothesis.  In  particular, with  deeper optical spectroscopy  of the
arc, our  redshift analysis  can be verified.   Moreover, observations
blueward  of the  blue channel  spectroscopy will  break  the redshift
degeneracy  of $z\!\simeq\!0.15$ and  $z\!\simeq\!2.5$.  Owing  to the
necessary  wavelength  ($\lambda_{\rm  obs}\!\lesssim\!3800$~\AA)  and
flux limit (${\rm  AB}\!\gtrsim\!26$~mag), these observations could be
conducted  with the  ultraviolet channel  of the  Wide  Field Camera~3
(WFC3)  for  HST.  Since  the  radio  source  has not  been  optically
identified, its  redshift can only be very  coarsely constrained.  The
majority of millijansky radio sources have early-type morphologies and
colors               at              $0.8\!\lesssim\!z\!\lesssim\!1.5$
\citep[eg.][]{win85,jqr07},  consequently  its Balmer/4000~\AA\  break
will     occur      at     $7000~{\rm     \AA}\!\lesssim\!\lambda_{\rm
obs}\!\lesssim\!1~\mu{\rm m}$.   Furthermore, \name\ is  undetected to
faint   flux-levels,   and   may   require  a   rather   deep   (${\rm
AB}\!\gtrsim\!25$~mag)  imaging campaign.   Therefore,  medium-band or
grism  observations  in  the  infrared  mode with  WFC3  would  better
constrain  the lens  redshift.  Should  such observations  support our
proposed lensing scenario, this system could be among the most distant
known                       gravitational                       lenses
\citep{castles}\footnote{\url{http://www.cfa.harvard.edu/castles/}}, a
dusty ERO with $M_*\!\lesssim\!10^{10}$~M$_{\odot}$, or a genuine dark
lens,  which  would  be  a  unique confirmation  of  the  $\Lambda$CDM
paradigm.

\acknowledgments   We  acknowledge   support  from   NASA   grant  HST
AR-10974.01A  (to RER), HST  AR-08357.01A (to  RAW) awarded  by STScI,
which is  operated by  AURA for NASA  under contract NAS  5-26555, and
NASA JWST  grant NAG5-12460  (to RAW).  We  thank Warren  Brown, Jacob
Russell, Rolf  Jansen, Alejandra Milone, and Nimish  Hathi for helpful
discussions  and  guidance.   We  are particularly  grateful  for  the
excellent comments and suggestions of the anonymous referee.

\end{document}